\newif\ifpreprint
\preprinttrue 

\newif\ifSupplementary
\Supplementaryfalse

\ifpreprint
\documentclass{natureprintstyle}
\else
\documentclass{nature}
\fi

\usepackage{epsfig}
\usepackage{color}
\usepackage{bm}
\usepackage{graphicx}
\usepackage{longtable}
\usepackage{amssymb}
\usepackage{tikz}
\usepackage{paralist}
\usepackage{lscape}
\usepackage[para,symbol]{footmisc}
\usepackage{aas_macros} 

\usepackage{amsmath}

\usepackage{soul}
\setstcolor{red}
\newcommand{\msolar} {$\rm{M_{\odot}}~$}
\newcommand{\msolarc} {$\rm{M_{\odot}}$}
\newcommand{\molH} {$\rm{H_2}$~}
\newcommand{\molHc} {$\rm{H_2}$}
\newcommand{\JJ} {\rm{$J_{21}$}~}
\newcommand{\JJc} {\rm{$J_{21}$}}
\newcommand{\zsolar} {$\rm{Z_{\odot}}~$}
\newcommand{\zsolarc} {$\rm{Z_{\odot}}$}
\renewcommand{\apj} {Astrophys. J.}
\renewcommand{\apjl} {Astrophys. J. Lett.}
\renewcommand{\apjs} {Astrophys. J. Suppl.}
\renewcommand{\mnras} {Mon. Not. R. Astron. Soc.}
\renewcommand{\nat} {Nature}
\renewcommand{\araa} {Ann. Rev. Astron. Astrophys.}
\renewcommand{\aap} {Astron. Astrophys.}

\def\textred#1{{\color{red} \bf #1}}

\def\figureOneCaption{
  \caption{\textbf{Modelling Synchronised Haloes.}  The synchronised proto-galaxy
          scenario. With only a 
          background field in operation a (delayed) Pop III star forms due to 
          \molH cooling (Case A). If a nearby star-burst galaxy, in conjunction with the 
          background, provides the critical LW flux required then a DCBH can 
          form in an atomic cooling halo (Case B). T$\rm{_{sync}}$ is defined as the time between
          the star-burst turning on and the point at which a PopIII would have formed. 
          T$\rm{_{on}}$ is the time taken for an atomic cooling halo to collapse and form 
          a DCBH (or the minimum time the source must shine for). \label{Cartoon}}
}

\def\figureOne{
    \setcounter{figure}{0}
    \ifpreprint \begin{figure} [h] 
        \centerline{
          \includegraphics[width=0.5\textwidth,trim=4pt 10pt 5pt 5pt, clip=false ]{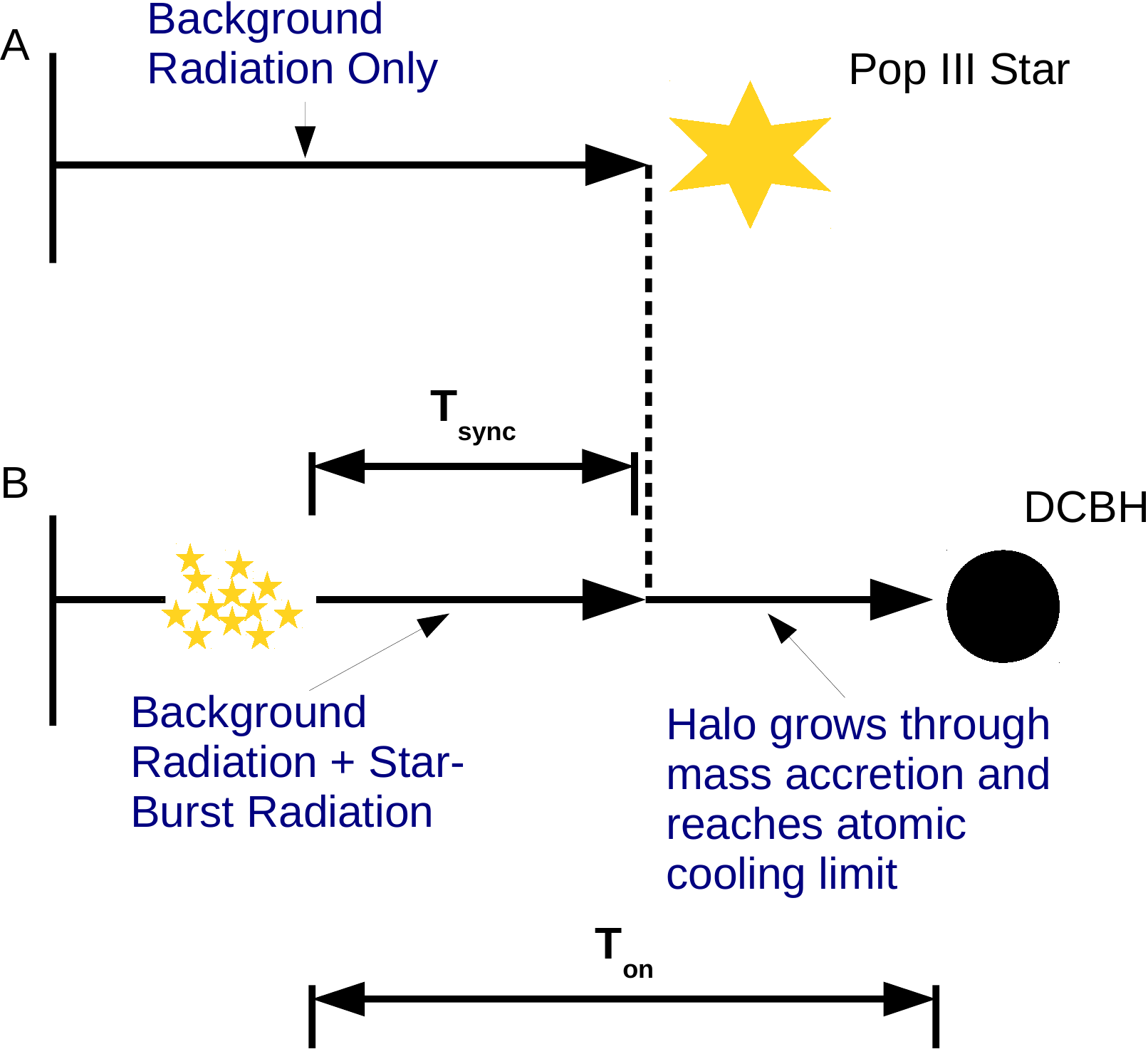}
        }
        \else \begin{figure} 
          \centerline{
            \includegraphics[width=0.85\textwidth, trim=4pt 10pt 5pt 5pt, clip=false]{Figure1.eps}
          }
           \fi
      \ifpreprint \figureOneCaption \else \caption{} \fi
        \ifpreprint \vspace{-4mm} \fi
        \end{figure}
}

\def\figureTwoCaption{
  \caption{\textbf{Ray Profiles for six selected haloes}.
          A selection of Ray Profiles from simulations using a star-bursting galaxy source as the
          local radiation source. For an atomic cooling halo to form the distance 
          to the source must, in general,  be less than approximately 300 pc. Four simulations
          make it to the atomic cooling track (green, black, cyan \& red lines) while two simulations
          (blue \& magenta) end up on the molecular cooling track. The simulations which ``failed''
          to achieve an isothermal collapse both show marked increases in \molH towards the
          very centre of the halo. The golden dashed line in the bottom right panel shows the
          level of the uniform background radiation employed (set at J$_{\rm BG} = 150$ \JJ in this
          illustrative case).\label{PopIII_Multiplot}}
}

\def\figureTwo{
    \setcounter{figure}{1}
    \ifpreprint \begin{figure} [h] 
        \centerline{
          \includegraphics[width=0.5\textwidth,trim=4pt 10pt 5pt 5pt, clip=false ]{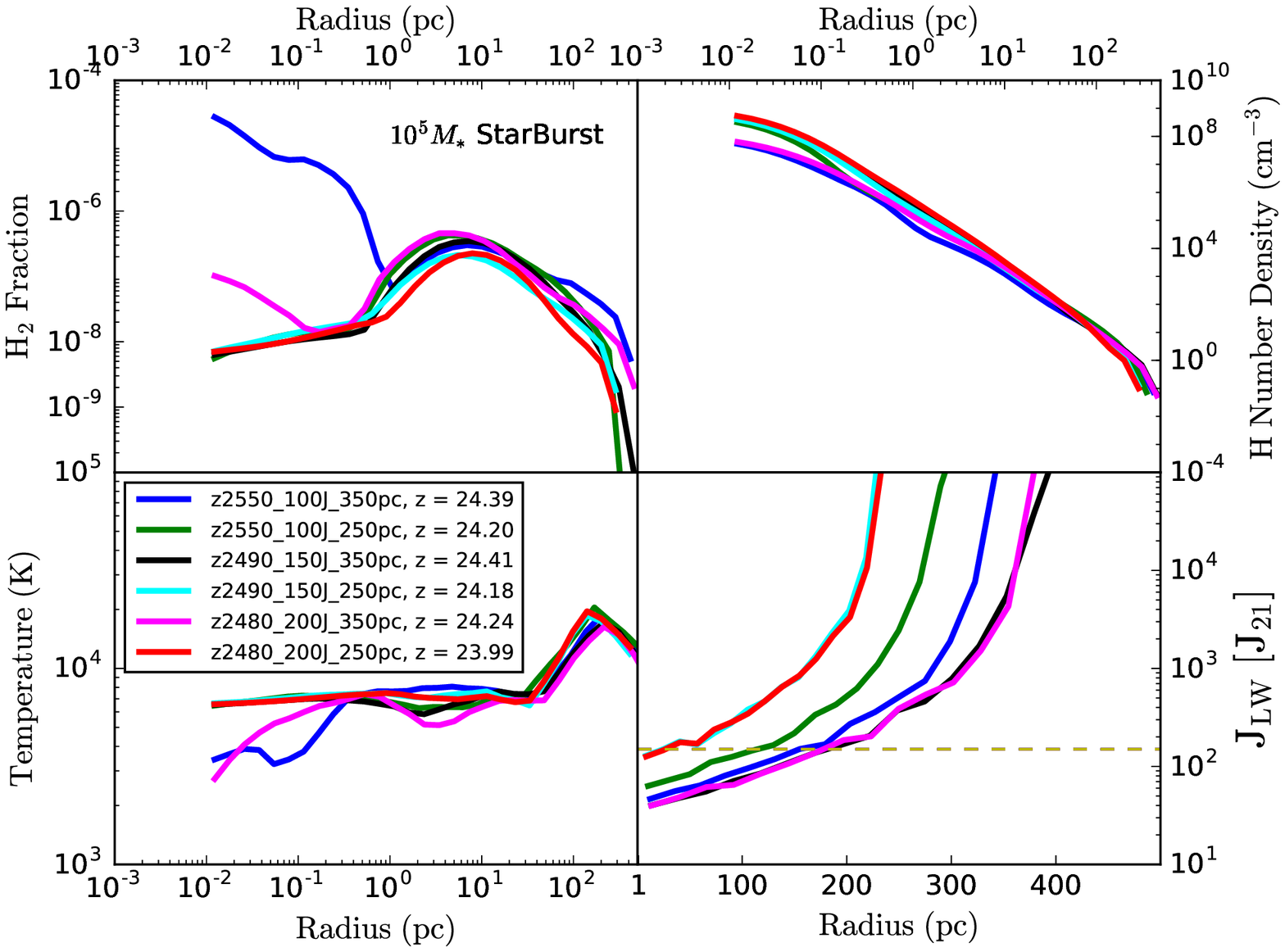}
        }
        \else \begin{figure} 
          \centerline{
            \includegraphics[width=0.85\textwidth, trim=4pt 10pt 5pt 5pt, clip=false]{Figure2.eps}
          }
           \fi
      \ifpreprint \figureTwoCaption \else \caption{} \fi
        \ifpreprint \vspace{-4mm} \fi
        \end{figure}
}

\def\figureThreeCaption{
  \caption{\textbf{The Synchronised Halo Zone.}
          This figure sums up the results of the suite of simulations conducted during this study. 
          Plotted on the x-axis is the synchronization time, and on the y-axis the average 
          separation between the source and the target halo. As the synchronization 
          time increases, the likelihood of attaining a DCBH diminishes as the 
          risk of metal pollution and photo-evaporation increases. Green crosses indicate cases where 
          an atomic cooling halo was achieved, orange crosses indicate cases 
          where an atomic cooling halo was achieved but the background was at our highest value
          of J$_{\rm BG}$ = 200 \JJc. Red crosses indicate simulations in which a
          molecular core formed in the centre of the collapsing halo. The green region
          in the centre is the region in which we find atomic haloes most likely to form.
          We also tentatively indicate (purple hatched region on the extreme left) a ``point of
          no return'' at which point we expect the formation of PopIII stars to be unstoppable
          independent of the intensity of the nearby start-burst.
          \label{Tsync}}
}

\def\figureThree{
    \setcounter{figure}{2}
    \ifpreprint \begin{figure} [ht] 
        \centerline{
          \includegraphics[width=0.5\textwidth,trim=4pt 10pt 5pt 5pt, clip=false ]{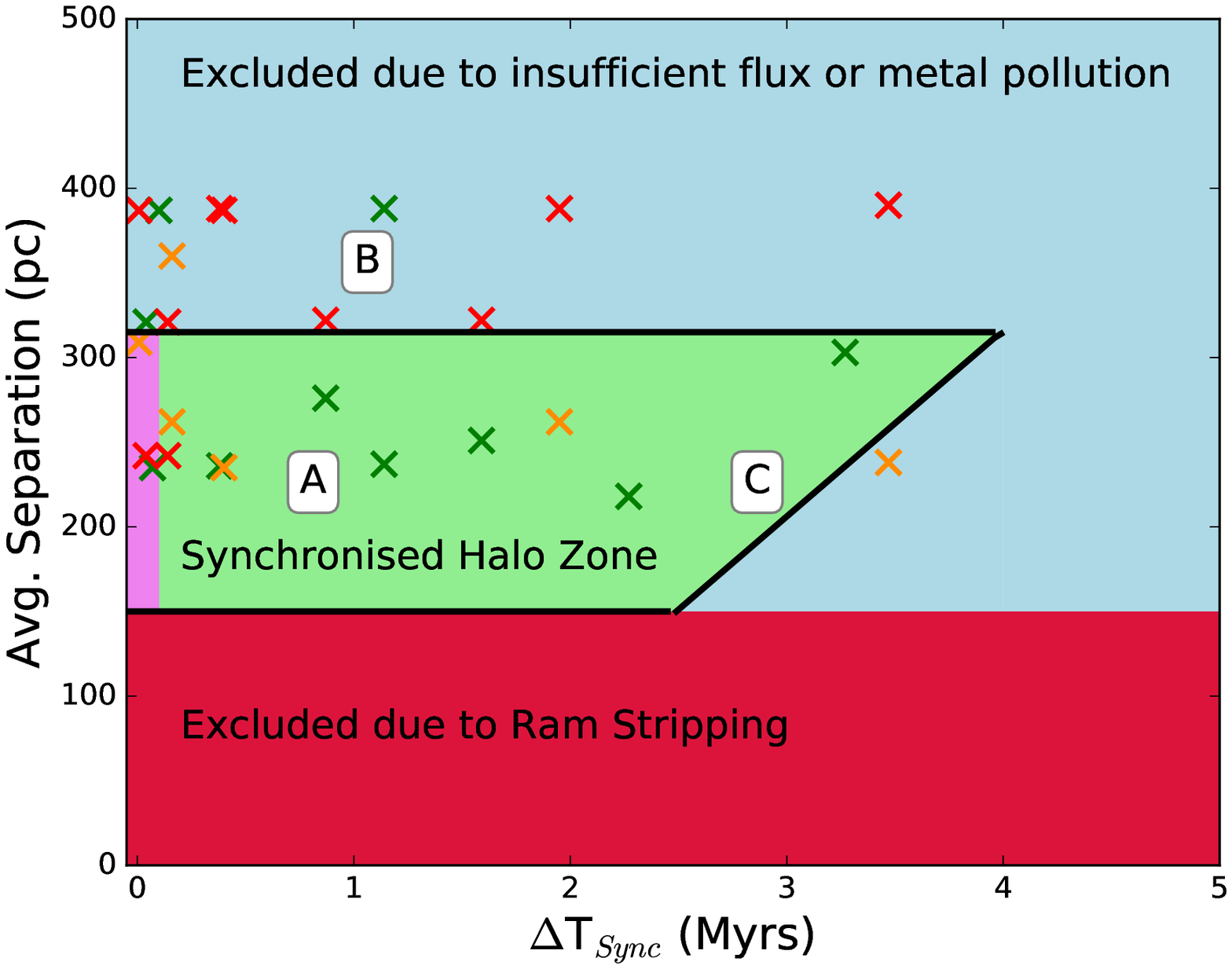}
        }
        \else \begin{figure} 
          \centerline{
            \includegraphics[width=0.85\textwidth, trim=4pt 10pt 5pt 5pt, clip=false]{Figure3.eps}
          }
           \fi
      \ifpreprint \figureThreeCaption \else \caption{} \fi
        \ifpreprint \vspace{-4mm} \fi
        \end{figure}
}

\def\figureFourCaption{
  \caption{\textbf{Mass Inflow Rates.}
          Mass Inflow Rates into the centre of collapsing halo for the same subset of simulations
          as shown in Figure 2.
          The mass inflow rates are calculated by determining the flux in spherical
          shells surrounding the central object. Focusing on the core of the collapsing halo
          where $R < 10$ pc, the mass inflow rates are typically much greater than 0.1 \msolar yr$^{-1}$
          with values peaking at around 0.75 \msolar yr$^{-1}$ at R $\sim 5 \times 10^{-2}$ pc.
          Interestingly, the highest mass inflow rates are seen for the simulations in which the flux
          is largest (i.e. the smallest separations). Finally, accretion rates greater than 0.1
          \msolar yr$^{-1}$ are sustained over several decades in radius. Below a scale of R $\approx
          3 \times 10^{-2}$ pc the inflow rates fall due to limited resolution effects below this
          scale. \label{AccRate}}
}

\def\figureFour{
    \setcounter{figure}{3}
    \ifpreprint \begin{figure} [h] 
        \centerline{
          \includegraphics[width=0.5\textwidth,trim=4pt 10pt 5pt 5pt, clip=false ]{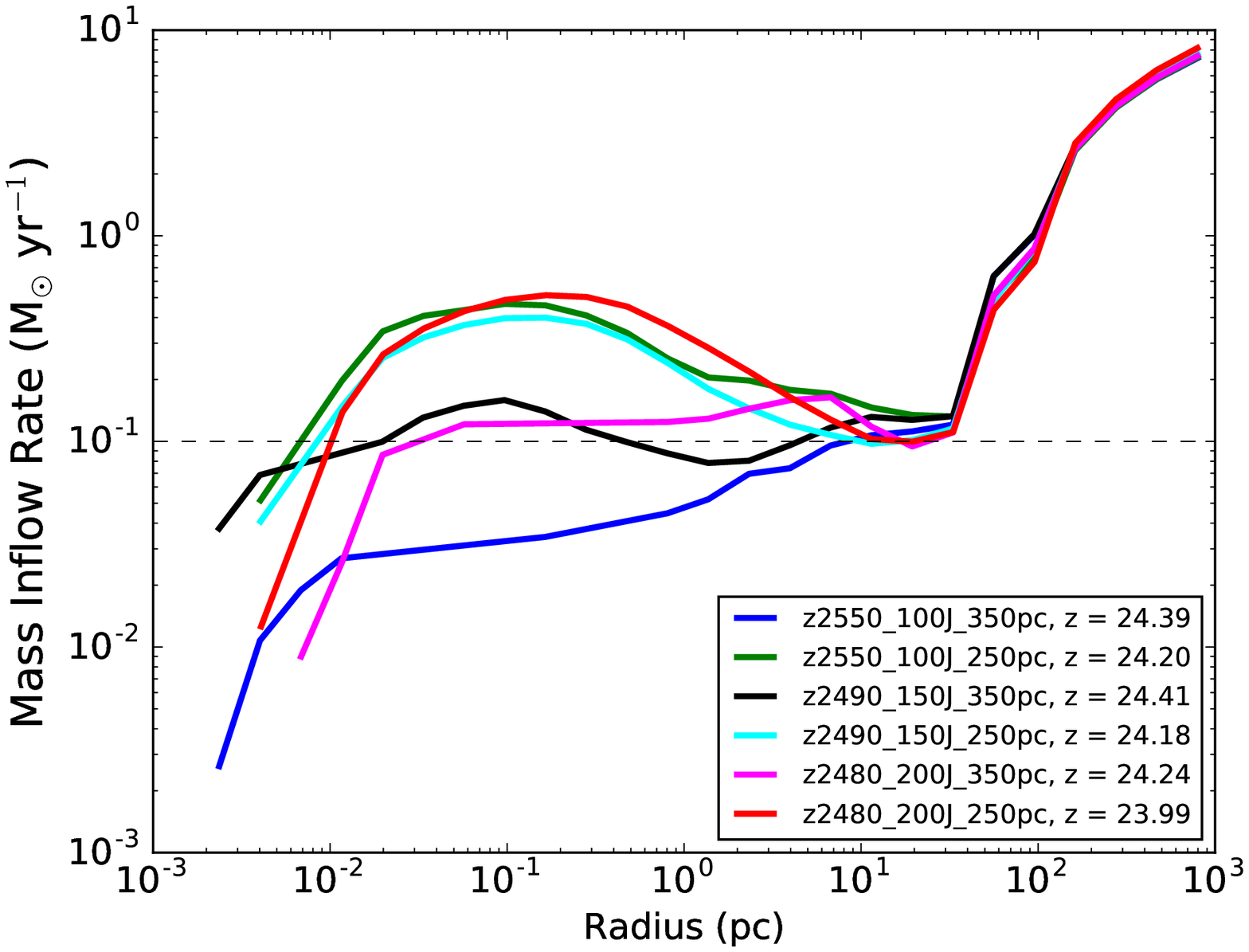}
        }
        \else \begin{figure} 
          \centerline{
            \includegraphics[width=0.85\textwidth, trim=4pt 10pt 5pt 5pt, clip=false]{Figure4.eps}
          }
           \fi
      \ifpreprint \figureFourCaption \else \caption{} \fi
        \ifpreprint \vspace{-4mm} \fi
        \end{figure}
}

\def\figureFiveCaption{
  \caption{  \textbf{Supplementary Data: Background Radiation Fields.}  Radial profile of
        the temperature for different background field strengths. The background field is
        modelled as a blackbody with an effective temperature of 30000 K. 
        The bright nearby source is not included in these simulations, and the radiation is
        purely that of the longer-lived background radiation field. The ``Ctrl'' field represents
        the no-background case, with the background field strength then increasing up to 1000 \JJc.
        For this halo realisation, the critical field strength is reached when the background field is
        $\sim$ 1000 \JJc. Achieving such a high background field in the high-z Universe is not viable,
        and in practice what will be required is a background augmented by an excess of local
        source(s), by unusually large streaming motions between baryons and dark matter, and/or
        by an unusually rapid merger history (see main text for discussion).\label{BackgroundFields}}
}

\def\figureFive{
    \setcounter{figure}{0}
    \ifpreprint \begin{figure*} [h]
    \textbf{\huge{Supplementary Information}}\par\bigskip\vspace{2cm}
        \centerline{
          \includegraphics[width=0.95\textwidth]{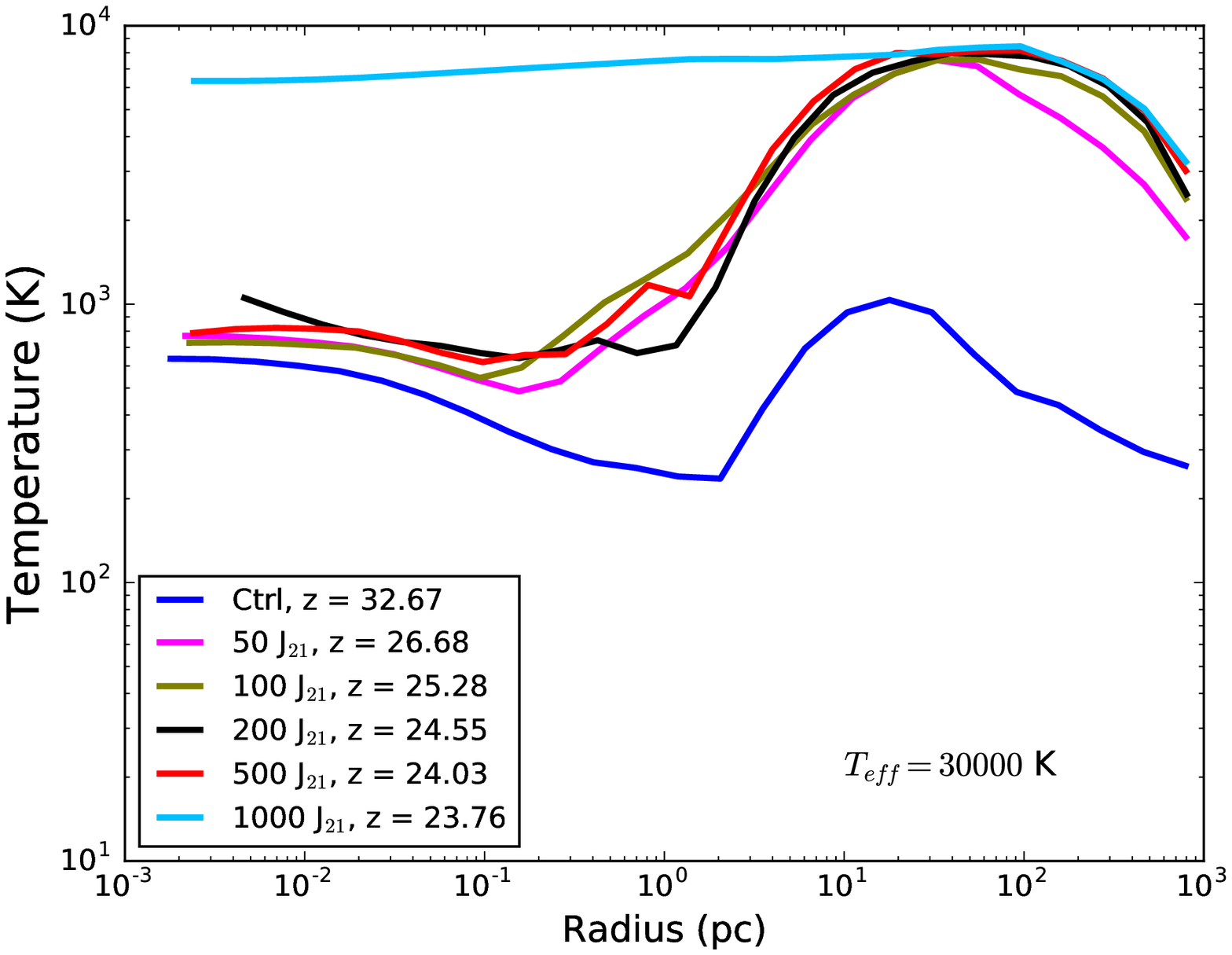}
        }
        \else \begin{figure} 
          \centerline{
            \includegraphics[width=1.0\textwidth, trim=4pt 10pt 5pt 5pt, clip=false]{BackgroundTemperatures.eps}
          }
           \fi
      \ifpreprint \figureFiveCaption \else \caption{} \fi
        \ifpreprint \vspace{-4mm} \fi
        \end{figure*}
}

\title{Rapid Formation of Massive Black Holes in close proximity to Embryonic Proto-Galaxies}

\author{$^*$John ~A. Regan$^{1,2}$, Eli Visbal$^{3,4}$, John ~H. Wise$^{5}$, Zolt\'an Haiman$^{3, 6}$,
    Peter H. Johansson$^{7}$ \&  Greg L. Bryan$^{3,4}$}

\begin{document}

\maketitle
\begin{affiliations}
 \item Institute for Computational Cosmology, Durham University, South Road, Durham, DH1 3LE, UK
 \item Centre for Astrophysics \& Relativity, School of Mathematical Sciences, Dublin City University, Glasnevin, D09 Y5N0, Dublin, Ireland
 \item Department of Astronomy, Columbia University, 550 West 120th Street, New York, NY, 10027, U.S.A.
 \item Center for Computational Astrophysics, Flatiron Institute, 162 5th Ave, New York, NY, 10003, U.S.A.
 \item Center for Relativistic Astrophysics, Georgia Institute of Technology, 837 State Street, Atlanta, 
   GA 30332, USA
   \item Department of Physics, New York University, New York, NY 10003, USA
\item Department of Physics, University of Helsinki, Gustaf H\"allstr\"omin katu 2a, FI-00014 Helsinki, Finland
\end{affiliations}
\begin{abstract}

\textbf{The Direct Collapse Black Hole (DCBH) scenario provides a solution for forming the massive
    black holes powering bright quasars observed in the early Universe. A prerequisite for forming
    a DCBH is that the formation of (much less massive) Population III stars be avoided - this can
    be achieved by destroying \molH via Lyman-Werner (LW) radiation (E$_{\rm{LW}}$ = 12.6 eV).
    We find that two conditions must be met in the proto-galaxy that will host the DCBH. First,
    prior star formation must be delayed; this can be achieved with
    a background LW flux of J$_{\rm BG} \gtrsim 100\ J_{21}$\footnote[2]{$J_{21}$ is the intensity of
    background radiation in units of $10^{-21} \rm ~erg ~cm^{-2} s^{-1} Hz^{-1} sr^{-1}$}.
  Second, an intense burst of LW radiation
  from a neighbouring star-bursting proto-galaxy is required, just before the gas cloud undergoes
  gravitational collapse, to finally suppress star formation completely. We show here for the
  first time using high-resolution hydrodynamical simulations, including full radiative transfer, that
  this low-level background, combined with tight synchronisation and irradiation of a secondary
  proto-galaxy by a primary proto-galaxy, inevitably moves the secondary proto-galaxy onto the
  isothermal atomic cooling track, without the deleterious effects of either photo-evaporating
  the gas or polluting it by heavy elements.  These, atomically cooled, massive
  proto-galaxies are expected to ultimately form a DCBH of mass $10^4 - 10^5 M_{\odot}$.}
\end{abstract}
The appearance of SMBHs at very early times in the Universe\cite{Fan_2006, Mortlock_2011, Wu_2015}
is a challenge to our understanding of star and black hole formation in the early Universe. Either
a select few Population III (PopIII) remnants must have undergone periods of prodigious
growth\cite{Madau_2001, Tanaka_2008, Madau_2014, Lupi_2016, Inayoshi_2016} or SMBHs could
  have been formed through massive galactic collisions\cite{Mayer_2010, Mayer_2014} or
alternatively SMBH seeds must have been massive ($\sim 10^{4}-10^{6}$ \msolarc) to begin
with\cite{Loeb_1994}. We here investigate the latter, so-called, DCBH scenario.
\ifpreprint \figureOne \fi
To probe the
unique combination of a background radiation field in tandem with an intense proximate burst we
perform a series of high resolution radiation-hydrodynamic simulations using the adaptive mesh
refinement code Enzo\cite{Enzo_2014} together with the Grackle multi-species library for solving
the primordial (H+He) chemistry network and regulating the radiation backgrounds\cite{Grackle}.
Radiative transfer is handled by Enzo's  MORAY ray tracing package\cite{WiseAbel_2011}.\\
\indent We employ relatively mild\footnote{relative to the rather high values
    often cited in the lierature where background fields of $\gg 1000$ \JJ are typically invoked}
backgrounds from between $\rm{J_{\rm BG} = 100}$ \JJ up to $\rm{J_{\rm BG} = 200}$ \JJ for an effective
background temperature of 30,000 K. This is the effective
temperature expected from a population of metal-free and partially metal-enriched stars 
in the early Universe. This (relatively mild) global background radiation field is sufficient to
delay the collapse but will not prevent the formation of \molH as the halo\footnote[3]{halo
  encompasses both the proto-galaxy and the dark matter structure surrounding the proto-galaxy}
mass increases.
To model the nearby source we use the  ``synchronised pairs'' scenario in which a proto-galaxy
(secondary halo) is exposed to the intense radiation from a star-burst (primary halo) that is
sufficiently near-by\cite{Dijkstra_2008, Yue_2014, Agarwal_2013, Agarwal_2014b} and tightly
synchronised in time\cite{Visbal_2014b}. The model is illustrated in Figure 1. If the primary halo
crosses the atomic cooling threshold (and begins forming stars through molecular collapse) closely
in time before a neighbouring, secondary halo, the primary halo can bombard the secondary halo with a
critical flux of LW radiation. This bombardment destroys the \molH in the secondary halo due to the
extremely high flux of the primary halo. It provides the
final push to the secondary halo forcing the halo onto the atomic cooling track leading to DCBH
formation. Forming massive black holes in this way is a promising way of producing massive black
hole seeds in the early Universe. \\
\ifpreprint \figureTwo \fi
\indent Further consideration must be given to preventing both metal contamination and
photo-evaporation (from photons with $\rm{E > 13.6}$ eV) from neighbouring haloes. Metal cooling will
rapidly reduce the temperature of the collapsing gas resulting in strong fragmentation (the cooling
time at a density of n$_{\rm HI}$ $\sim$ 100 cm$^{-3}$ is about 10 Myr for gas with a metallicity of
$Z=10^{-4}-10^{-3}$ \zsolarc ). Therefore, for the Jeans mass to remain large, metal pollution from
nearby galaxies must be avoided (although this can be mitigated by inefficient or slow mixing due
to external metal enrichment\cite{Smith_2015}). In addition, photo-evaporation in haloes exposed
to star-burst radiation for longer than $\sim 40$ Myrs from a source with luminosity
L = $1.64 \times 10^{41}$ erg/s and a separation of between 0.5 and 1 kpc was observed in the
simulations of Regan et al. (2016)\cite{Regan_2016a}, effectively limiting the time available
for creating a pristine atomic cooling halo (ACH) using a nearby neighbour. \\
\indent Our results show that in order to achieve an ACH the separation, $\rm{R_{sep}}$, between the
primary and secondary haloes needs to be less than approximately 300 pc for the stellar luminosity
of the primary halo adopted in our model ($1.2 \times 10^{52}$ photons/sec, or a stellar mass of
$\rm{M_{*} \sim 10^5\,M_\odot}$). This critical distance will vary as $\rm{d_{crit} \propto
  \sqrt{M_{*}}}$ but for the luminosities examined in this paper will be approximately 300 pc. 
We define the synchronisation time as the time between when the primary halo is turned on and when
the gas in the secondary halo would have collapsed in the absence of the primary halo - see Figure 1.
The synchronisation times that successfully produce an ACH can range from  a few kyr to a few Myr
(see column 6 in Supplementary Data Table 1). Short synchronisation times reduce the probability
of metal pollution and/or photo-evaporation from the primary halo. \\
\indent In Figure 2  we show a selection of Ray Profiles using our primary halo as the local
galactic source (the stellar mass of the primary halo was determined from high resolution simulations
of the high redshift Universe\cite{Xu_2013}). For each simulation, we show the temperature,
\molH fraction, hydrogen number density and the intensity in the LW band as a function of 
distance from the center of the secondary halo. Of the six cases examined
in Figure 2 four collapse nearly isothermally and form ACHs, while the two others form excessive
\molHc, cooling significantly.
Three of the ``successful'' simulations lie within the critical distance of 300 pc, with the fourth 
at 388 pc. The thermal profile of the fourth simulation is not as smooth as the closer-in 
simulations and this case is therefore likely close to a tipping point (i.e. a slightly larger 
separation would likely have resulted in a non-isothermal collapse).  The virial radius of our 
secondary halo is approximately 300 pc - meaning that for the neighbouring source to have the 
greatest probability of completely suppressing \molH cooling the virial radii of the primary and
secondary must overlap\cite{Visbal_2014b}. Primary sources which lie 
outside of the virial radius of the secondary do not, in general, provide sufficient flux without 
unrealistically high star-formation efficiencies. \\

\noindent \textit{\textbf{How spatially and temporally correlated do the halo pairs need to be?}} \\
\ifpreprint \figureThree \fi
In Figure 3 we show the average separation against the synchronisation time, $\rm{T_{sync}}$,
for all of the simulations conducted in this study. As noted above, for separations of
$\rm{R_{sep}}\gtrsim 300 \ \rm pc$ we tend not to form ACHs and instead a critical
level of \molH builds up leading to a non-isothermal collapse.
Due to the possibility of metals from supernova explosions polluting the pristine environment of the
secondary halo and the detrimental effects of long term exposure of the gas to ionising radiation
we disfavour sources which must be ``on'' for more than 10 Myrs (which is comparable to the lifetime 
of massive stars) before the initial collapse occurs. We therefore do not probe backgrounds
with $J_{BG} < 100$ \JJc. As a result of these constraints, regions at the top and to the right in
Figure 3 are excluded. Ram pressure stripping will affect haloes which get too close
to one another and thus the bottom section of the graph is excluded for this
reason\cite{Visbal_2014b, Chon_2016} (see Methods section for further discussion). 
This leaves a (green) region in the left centre of the Figure which allows for the formation of
DCBHs. The crosses are results from the suite of simulations (see Supplementary Data Table 1). 
The simulations which formed an atomic cooling halo are labelled with green crosses 
and live in this region. Orange crosses indicate simulations which formed an atomic cooling halo
but for which the background was set to J$_{\rm BG}$ = 200 \JJc (the maximum level in our study).
This is quite a strong background and may be beyond even the most clustered regions\cite{Ahn_2009}.
Red crosses indicate simulations in which a non-isothermal collapse was observed due to \molH
cooling and in which a DCBH is therefore unlikely to form.\\
\indent Three regions of particular interest are marked on the plot as A, B \& C.
Region A is where most of our simulation results cluster. Short (T$\rm{_{sync}} \lesssim 2$ Myr) 
synchronisation times combined with close separations almost always resulted in an ACH.
Region B is outside our ``Synchronised Halo Zone'' (SHZ) and here the majority of the simulations
show that the flux received by the secondary is too small and an atomic cooling halo does not form.
Nonetheless four atomic cooling haloes appear outside the SHZ. Two of the green crosses are from
runs with a background of J$_{\rm BG}$ = 150 \JJ while one is from a run with a background
of J$_{\rm BG}$ = 100 \JJc. The orange cross had a background of J$_{\rm BG}$ = 200 \JJc. Similarly,
two molecular collapses are seen inside the SHZ showing that 
fluctuations limit the accuracy of the SHZ boundary at approximately the 10\% level. Region C 
indicates the tip of the SHZ. As the synchonisation time gets longer the risk of photoionisation 
and/or metal pollution becomes larger.
Synchronisation limits will vary depending on the exact environmental conditions, the luminosity,
the spectrum and the distance to the primary halo, we 
therefore only probed the limits of the separation and the limits of the synchronisation 
(i.e. how short/extended can the synchronisation time be). \\
\indent Forming seeds for the SMBHs we observe at the centres of the most massive galaxies and as
very high redshift quasars is an outstanding problem in modern astrophysics. We show here that the
combination of a relatively mild ``local'' radiation background field due to the clustering of early galaxies
plus a nearby star-burst event is the perfect trigger for the creation of such an atomic cooling
halo. The local background serves to delay PopIII star formation allowing
a sufficiently massive halo to develop. A nearby (synchronised) star-burst can then
irradiate the now massive halo with a flux greater than the critical LW flux pushing the collapsing
halo onto the atomic cooling track while avoiding the deleterious effects of photo-evaporation
or metal pollution. We find that a ``synchronised halo zone'' exists where the separation between the
neighbouring haloes is between approximately 200 pc and 300 pc and the synchronisation time between
the halo's evolution is less than approximately 4 Myrs (``on'' time $< 10$ Myrs).  Furthermore,
we find the mass inflow rates onto the central object are greater than 0.1 \msolar yr$^{-1}$
over several decades in radius. \\
\ifpreprint \figureFour \fi
\indent Close halo pairs
with tight synchronisation times should easily fulfill the number density requirements of SMBHs at
high redshift\cite{Dijkstra_2008, Ahn_2009, Visbal_2014b} (see methods section for further
  details and a calculation of expected number densities) and may play a significant role in the
formation of all SMBHs. Upcoming observations with the JWST will search for DCBH candidates up to
redshifts of $z\sim 25$, and may be able to identify them based on their unique spectral
signatures\cite{Pacucci_2015, Natarajan_2016}. A detection of a DCBH candidate together with
star-forming galaxies in close proximity would validate the synchronisation mechanism.

\begin{methods} \label{Methods}

  \subsection* {\\Simulation Setup.\\}
We use the adaptive mesh refinement code Enzo in this study. Enzo uses an N-body adaptive
particle-mesh solver to follow the dark matter dynamics. It solves the hydrodynamics equations
using the second-order accurate piecewise parabolic method, while an HLLC Riemann solver ensures
accurate shock capturing with minimal viscosity. Rather that using the internal chemistry solver we
use a modified version of the Grackle chemistry library which has been updated with the latest
rates for modelling collapse in the face of radiation backgrounds\cite{Glover_2015a, Glover_2015b}.
The chemical network includes 33 separate chemical reactions (see Table 1 from \cite{Regan_2016a})
from 10 species -
${\rm H}, {\rm H}^+, {\rm He}, {\rm He}^+,  {\rm He}^{++}, {\rm e}^-,$ $\rm{H_2}, \rm{H_2^+},
\rm{H^-} \rm{and}\ \rm{HeH^+}$. \\
\indent The initial conditions are taken from Regan et al. (2015)\cite{Regan_2015} and were generated
with the MUSIC initial conditions generator\cite{Hahn_2011}. Haloes were initially located by running
a large suite of dark matter only simulations in a cosmological context. Realisations were then
selected by choosing simulations in which massive dark matter haloes formed relatively early and thus
represented high ``sigma'' peaks in the primordial density field. Simulations were then rerun with
gas dynamics included. All simulations were run within a box of size 2 $h^{-1}$ Mpc (comoving), the
root grid size was $256^3$ and we employed three levels of nested grids. Within the most refined
nested grid the dark matter resolution was M$\rm{_{DM}} \sim 103$ \msolarc. In order to increase further
the dark matter resolution we split the dark matter particles\cite{Kitsionas_2002} at a redshift of
z = 40, well before the onset of the collapse. This has no adverse effect on the dynamics of the
collapse but is a necessary step in high resolution simulations\cite{Regan_2015}. The baryon
resolution is set by the size of the highest resolution cells within the grid. Grids are refined
in Enzo whenever, user defined, criteria are breached. We allow refinement of grid cells based on
three physical measurements: (1) the dark matter particle overdensity, (2) the baryon overdensity and
(3) the Jeans length. The first two criteria introduce additional meshes when the overdensity
$\big( {\delta \rho \over \rho_{mean}} \Big)$ of a grid cell with respect to the mean density exceeds
8.0 for baryons and/or dark matter. Furthermore, we set the \textit{MinimumMassForRefinementExponent}
parameter to -0.1 making the simulation super-Lagrangian and therefore reducing the threshold for
refinement as higher densities are reached. For the final criteria, we set the number of cells per
Jeans length to be 16 in these simulations. We set the maximum allowed refinement level to 18. This
means that we reach a maximum grid resolution of $\Delta x \sim 2 \times 10^{-3}$ pc
(physical at z = 25).\\

\subsection*{Choosing a background radiation field.\\} \label{bfield_choosing}
The vast majority of studies undertaken to investigate the destruction of \molH by an
external radiation field have used a uniform background radiation field to demonstrate the
viability of the mechanism. The general consensus has been that a critical LW intensity of
$\rm{J \gtrsim 1000}$ \JJ
is needed\cite{Shang_2010, Latif_2014a, Latif_2014b, Agarwal_2015a} for a blackbody background with
an effective temperature of $10^5$
K\footnote[3]{A background intensity approximately an order of magnitude lower would be sufficient
  for low-mass stars with an effective temperature of $T = 10^4 K$ \cite{Omukai_2001, Shang_2010};
  however the stellar mass required is then significantly
  higher \cite{WolcottGreen_2012,Wolcott-Green_2016}.}.
However, achieving a background of $\rm{J_{BG} \gtrsim 1000}$ \JJ is very unlikely at the very high
redshifts of interest ($z\gtrsim 25$)\cite{Ahn_2009}. Instead what is much more likely to create
the required flux is a mild background ($\rm{J_{BG} \sim 100 - 200}$ \JJ) augmented by a local burst
of LW radiation from one or more
nearby sources\cite{Dijkstra_2008, Dijkstra_2014, Visbal_2014b}. The synchronised pairs scenario
requires a mechanism to delay the formation of PopIII stars so that neighbouring haloes can build
sufficient mass to cross the atomic cooling threshold in near
synchronisation. In this study we impose an external background field to facilitate this
delay\cite{Machacek_2001, OShea_2008}. For the halo studied in this work we found, in agreement
with other studies in the literature, that
a background radiation field of $\sim$ 1000 \JJ is required to form an atomic halo with a blackbody
effective temperature of 30000 K (when \textit{only} a global background is considered). 
In Extended Data Figure 1 we have plotted the temperature as a function of radius for six different
background fields. This figure contains no local source and shows the impact of the background
radiation field only. The ``Ctrl'' field has no background radiation field invoked and so the halo
collapses earliest in this case. As the strength of the radiation field is increased the collapse
is delayed to increasingly longer times until eventually the collapse becomes
isothermal and the halo collapses under the influence of atomic hydrogen cooling only.  \\
\indent We refer to these background fields as ``local backgrounds''. These are intensities
produced by a collection of nearby galaxies which produce radiation which locally
acts like a uniform background. The local background delays PopIII star formation in the secondary halo
of our simulations. This is crucially important in the context of exploring the temporal elements
of our model. If the local background is too weak then the conditions for halo synchronisation
cannot be met (although streaming motions and/or rapid collapse, which are not included in our
simulations, could obviate the need for the background; see below).  In principle, the star-burst
radiation could compensate for the lower background field. This requires, however, the star-burst
to shine for a longer time, increasing the risk of metal pollution or photo-evaporation. In
this study we examine background radiation fields, J$_{\rm BG}$, with values of 100 \JJ, 150 \JJ
and 200 \JJc. In all cases we initiate our background radiation field at z = 35.\\

\subsection*{Radiation Source.\\} \label{source}
In order to study the effect a local radiation source (i.e. the primary halo) can have on the 
``target'' halo (i.e. the secondary halo) we employ a massless radiation particle for which
we can vary the emission intensity and source separation. The radiation particle acts as
  a source of radiation, the radiation is propagated by Enzo's radiative transfer
  scheme\cite{WiseAbel_2011} which traces individual photon packets through the AMR mesh by
  utilising 64 bit HEALPix algorithms\cite{AbelWandelt_2002, Gorski_2005}. The ray tracing scheme
  allows us to follow photons in the infrared,
  LW and hydrogen ionisation range. The exact energy levels and relative strength of each energy
  bin used are given in Supplementary Data Table 2. By utilising
  the ray tracing scheme we are able to accurately track both the column density along each ray
  and the level of photodissociation of different ions along the ray path. The shielding effects
  of different species are accounted for by the exact knowledge of the column density (and therefore
  the optical depth). Self-shielding of \molH is accounted for using the prescription given in
  Wolcott-Green et al. (2011)\cite{Wolcott-Green_2011}. \\
\indent We control the physical distance between the
primary and the secondary haloes by first running a test simulation where the radiation particle is
placed at a distance of approximately 200 pc from the source. The simulation is then run and we
calculate the position of the point of maximum density averaged over multiple outputs. By doing
this we know in advance how the centre of mass (COM) of the system will change as the simulation
proceeds (as the initial source position is changed the COM will change but our numerical
experiments showed that the change was only at the 10\% level). We then choose a vector with the
COM as the origin. By then placing the radiation particle at set distances along this given vector
(we use the angular momentum vector for convenience) we find points where the distance to the
secondary halo remains approximately constant - although some variation is expected.  This mimics
a scenario where the secondary halo and primary halo orbit one another. \\ 
\indent The radiation spectrum emitted by the primary halo is modelled to be a partially metal 
enriched galaxy. We use a luminosity of $\sim 1.2 \times 10^{52}$ photons per second 
(above the H$^-$ photo-detachment energy of 0.76 eV - see Figure 2 in Regan et al. 2016a). 
We assume a stellar mass of $M_*=10^5$ \msolar at $z=20$, consistent with the largest galaxies 
prior to reionisation in the Renaissance Simulations\cite{Chen_2014}.  We then calculate its 
spectrum using the Bruzual \& Charlot (2003) models\cite{Bruzual_2003} with a metallicity of 
$10^{-2}$ \zsolar and compute the photon luminosity from it. The spectrum does not include
emission from the nebular component and is solely due to stellar emission.  This spectrum is 
virtually identical to a PopIII source with a cluster of 9 \msolar stars\cite{Schaerer_2002}. 
However, a cluster of $10^4$ PopIII stars is a challenging structure to produce even under 
extreme conditions\cite{Chen_2014} and hence we opt for the metal-enriched population. The details 
of the spectrum (and a comparable PopIII spectrum for comparison) is detailed in Supplementary
Data Table 2.\\

\subsection*{Measuring the Radiation Intensity for a Local Source.\\} 
J$_{21}$ is the standard unit used to measure radiation (background) intensities. 
To calculate and quote an equivalent angle-averaged intensity from a single nearby source, 
$J_{\rm L}$, in units of J$_{21}$ we sum the contributions from radiation in the LW band only.
It should be noted however that the radiation intensity includes effects from radiation at other 
energies also. 
\begin{align}
J^\prime & \equiv  {k_{\rm H2I}  E_{\rm LW} \over 4 \pi \sigma_{\rm H2I}(E_{\rm LW}) } \\
J_{\rm L}  & \equiv {J^\prime \over \nu_{\rm H} J_{21}}
\end{align}
where $J^\prime$ is the intensity in the LW band. Here $k_{\rm H2I}$ is the number of 
photo-dissociations per second for \molH, $\sigma_{\rm H2I}(E)$ is the average cross section for 
dissociation in the LW band\cite{Abel_1997} and E$_{\rm LW}$ is the photon energy in the LW band.
Finally, $\nu_{\rm{H}}$ is the frequency at the hydrogen ionisation edge. $J_{\rm L}$ is now the intensity
of the radiation from the local source in units of J$_{21}$ normalised at the Hydrogen ionisation
edge. $J_{\rm L}$ will be a function of distance from the source i.e. $J_{\rm L}$ = $J_{\rm L}(d)$. \\

\subsection*{Ram Pressure Stripping \& Tidal Disruption.\\}
The small separation of the primary and secondary haloes implies that 
they are sub-haloes of a larger parent halo. As they move through the ``intra-cluster medium''
of the parent halo, they are subject to ram pressure stripping. The ram pressure due to this 
movement is:
\begin{equation}
P_{\rm ram} = \frac{1}{2} \rho_{\rm m} v^2.
\end{equation}
To ensure that this ram pressure does not unbind the core of the
secondary galaxy, P$_{\rm ram}$ must obey\cite{Gunn_Gott_1972, Gisler_1976,
McCarthy_2008} 
\begin{equation}
P_{\rm ram} \le {\alpha\ G \ M_{\rm tot} (R_{\rm core}) \ \rho_{\rm gas} (R_{\rm core}) \over R_{\rm core}} \label{Eqn:Ram},
\end{equation}
where $\rho_{\rm m}$ is the gas density the core passes through, $v$ is the relative orbital velocity 
of the proto-galaxy, $\alpha$ is a variable of order unity which depends on the specifics of the 
dark matter and gas profiles, $\rm{ M_{tot} (R_{core})}$ and  $\rm{\rho_{gas} (R_{core})}$ are the total
mass and gas density within $\rm{R_{core}}$ respectively. For the secondary halo studied here we
define the core of the halo as the radius at which the gas mass dominates over the dark matter mass
giving us a value of $\rm{R_{core}} \sim 10$ pc. Using the virial mass of the secondary halo we
assume a relative orbital velocity of 20 km/s, we then use the gas density at an orbital distance
of 100 pc ($\rm{\rho_m(R = 100\ pc) \approx 1 \times 10^{-23}\ g\ cm^{-3}}$) from the centre of the
secondary halo to get 
\begin{equation}
P_{\rm ram} \approx 2 \times 10^{-8} \rm{\ g\ cm^{-1}\ s^{-2}}
\end{equation} 
while the right hand side of equation \ref{Eqn:Ram} gives a value of 
\begin{equation}
{\alpha G M_{\rm tot} (R_{\rm core}) \rho_{\rm gas} (R_{\rm core}) \over R_{\rm core}} \approx 2 \times 10^{-8} \rm{\ g\ cm^{-1}\ s^{-2}},
\end{equation}
at a radius of 10 pc. Inside of this radius the binding energy of the halo increases
approximately as R$^{-2}$ and so within 10 pc ram pressure stripping will be unable to unbind the gas. 
To be conservative we always constrain the radiation particle to be at a distance greater than 150 pc
from the secondary halo centre. However, while the core will remain intact some mass loss
  from the outer parts of the protogalaxy is inevitable. Given the secondary and primary haloes
  are expected to eventually merge the total mass available for the accreting DCBH is unaffected. \\

\subsection*{Metal Pollution, Photo-evaporation and the Maximum Irradiation Time.\\} \label{pollution}
Metal pollution, photo-evaporation or the natural end of the star-burst in the primary galaxy will
ultimately limit the prospects for achieving a direct collapse black hole. A PopIII source model
with individual stellar masses of M$_{*} \sim 9$ \msolar will have an expected lifetime of 
$\rm{T_{*} \sim}$ 20 Myrs\cite{Schaerer_2002}. After this time the stars would explode as supernovae
and presumably pollute the secondary halo within a few Myrs\cite{Dijkstra_2014}. Our galaxy model
includes lower mass stars with longer lifetimes but with a more distributed IMF and so also stars with
masses in excess of 9 \msolarc. To be conservative we assume that in cases where the primary
halo must be ``on'' for greater than 10 Myrs that a DCBH does not form. We note that we see no
evidence of photo-evaporation of the secondary halo even when the secondary is irradiated for 22 Myr
(the longest ``on'' time we encountered during our numerical experimentation) but since we do not
include the impact of metal production and pollution we limit the maximum irradiation time to 10
Myrs.  \\
\indent A related issue concerning metal pollution is the impact that the background galaxies,
which currently are supposed to provide the background radiation field, may have. Dijkstra et al.
(2014)\cite{Dijkstra_2014} show that the metal pollution radius scales as
$\rm{r_s (M,t) \propto M_*^{1/5} t^{2/5}}$. If we assume that the radiation field is created by a
collection of 5 galaxies of similar mass to our primary galaxy then the metallicity field spreads as 
$\rm{r_{kpc, s} = 3 \times 10^{-2} \big( 5 \times 10^5\ M_{\odot} \big)^{1/5} n^{-1/5} t_{6}^{2/5}}$
(Dijkstra et al. (2014) equation 5) this leads to a spread of just over 2 kpc after 30 Myrs. This
suggests that for background sources separated by a distance greater than 2 kpc from the primary
and secondary haloes that metal pollution may not be such a large concern at the redshifts explored
here. \\
\indent A recent study by Chon et al. (2016)\cite{Chon_2016} examined the formation of DCBHs in a
large cosmological simulation. The focus of their simulations was to determine the impact of
tidal disruption and ram pressure on the formation of DCBHs. 
In a study of 42 DC candidates they found that two candidates successfully resulted in the rapid
formation of stars within an ACH. One of the two successful candidates was due to a scenario similar
to the synchronised scenario investigated here. However, their simulations were unable to
accurately track the radiation coming from the primary halo and were not focused on this scenario. 
In their case they found that the primary halo exceeded the atomic
cooling threshold 20 - 30 Myrs before the secondary halo with the LW intensity exceeding the
critical value. This suggests that we are quite conservative with our 10 Myr limit on the primary
halo ``on'' time. However, similar to this study, Chon et al. do not include the effects of metal
pollution which could limit
the time for which the star-burst can be active before the secondary must collapse under atomic
cooling.  We also note that Chon et al. did not include the impact of photoionisation in
their hydrodynamics simulations, whereas we do so here, and are therefore able to show that
photo-evaporation is avoided.\\

\subsection*{Mass In-Flow Rates. \\} \label{mass_inflow}
Large mass inflow rates are one of the pre-requisites for forming a supermassive star (SMS).
For the case of (super) massive stars the gravitational contraction timescale is much shorter
than the Kelvin-Helmholtz timescale meaning that the star begins nuclear burning before it has
finished accreting and in fact will continue to accrete (subject to the correct environmental
conditions) throughout its lifetime. Its final mass then becomes the important quantity. \\
\indent In Figure 4 we have plotted the mass inflow rates as found for the same subset of simulations
shown in Figure 2. The mass inflow rates are calculated in spherical shells around the central
density according to the equation 
\begin{equation}
\dot{M}(t) = 4 \pi R^2 \rho (R) V(R)
\end{equation}
\noindent where $\dot{M}(t)$ is the mass inflow rate, R is the radius, $\rho$ is the density and 
V(R) is the radial velocity at R. We find that the mass inflow rates peak at approximately
0.75 \msolar yr$^{-1}$ at R $\sim 5 \times 10^{-2}$ pc. Most importantly, mass inflow rates greater
than 0.1 \msolar yr$^{-1}$ are sustained over several decades in radius. Furthermore, we find that the
more intense the flux (i.e. the closer the separation) the higher the mass inflow rates. This
feature is likely due to the higher LW flux experienced at this radius, which should in turn lead
to high gas temperatures due to the higher dissociation rates of \molHc. \\
\indent The classical assumption regarding the formation of SMSs is that accretion rates
greater than at 0.03 \msolar yr$^{-1}$ are required\cite{Hosokawa_2012, Hosokawa_2013}. At this
mass accretion rate the protostellar evolution changes completely. The stellar envelope swells
greatly in radius reaching up to 100 AU. The effective stellar temperature drops to close to 5000 K
meaning that radiative feedback from the protostar becomes ineffective at preventing continued
accretion. In a somewhat complementary scenario a ``quasi-star'' may be born when high accretion rates
($\dot{M}(t) \gg 0.14$ \msolar yr$^{-1}$)\cite{Schleicher_2013} onto an already existing SMS star
results in the collapse of the hydrogen core into a stellar mass
sized black hole. The highly optically thick gas which keeps falling onto the black hole bringing with
it angular momentum results in an accretion disk forming around the central black hole. The energy
feedback inflates the innermost part of the inflow resulting in a ``quasi-star''. The mass loss
rates from these two types of objects is an area of active research with possible negative
conquences in the case of quasi-stars\cite{Fiacconi_2016} which may be absent in the case of
more ``normal'' SMSs\cite{Nakauchi_2016}. Either way the accretion rates observed in our simulations
satisfy the basic requirements that accretion rates of $\gtrsim$ 0.1 \msolar yr$^{-1}$ are
found over several decades in radius. \\
\indent It should be noted than below a radius of approximately
$3 \times 10^{-2}$ pc the inflow rates shown in Figure 4 fall. This is a purely numerical effect created by
both a diminising resolution below that scale (the Jeans mass of the gas at this temperature and
density is approximately 0.1 pc) and the fact that we stop the calculation once the gas reaches
our maximum refinement level. We are therefore not evolving the gas at this scale over any
significant fraction of its dynamical time. Further exploration of the inflow rate at this scale
would require the adoption of sink particles.\\
\indent Finally, it is not certain that if a monolithic inflow of $> 0.1$ \msolar yr$^{-1}$ is
attained and then sustained over an extended period of up to 1 Myr (possibly up to 10 Myrs as
required for quasi stars) that a single central object will form. Fragmentation may in this case
still occur within the collapsing object\cite{Regan_2014a} or else within a self gravitating disk
around the central object\cite{Inayoshi_2014b}. However, in either case the fragments are likely
to form a dense cluster and ultimately a DCBH\cite{Regan_2009b}.\\

\subsection*{Expected and Required LW Background.\\} \label{bfield_analysis}
Our simulations indicate that a background LW intensity of $J_{\rm BG} \gtrsim 100~J_{21}$ is 
required for DCBH formation in the synchronised haloes scenario. Determining the precise
abundance of DCBHs formed as a result of this requirement is beyond the scope of the present work, 
however, we argue here that such a high flux can plausibly be achieved and can potentially result
in a DCBH number density large enough to explain observations of SMBHs at $z \gtrsim 6$. \\
\indent The intensity of the LW background at $z \gtrsim 10$ is highly uncertain. Some theoretical
estimates have found mean values between $(0.1-1) J_{21}$ at $z=25$ and roughly an order of
magnitude higher values by $z=10$\cite{2013MNRAS.432.2909F, 2014MNRAS.445..107V, 2015MNRAS.453.4456V}.
Even though the $z=25$ values are 2-3 orders of magnitudes below the requirement found 
in our simulations, such a high background may still be possible in overdense environments.
The simulations of Ahn et al. (2009)\cite{Ahn_2009} show that even in a relatively small box
(35 ${\rm Mpc}~h^{-1}$) there are regions with LW flux several orders of magnitude higher than the
minimum value (see their Figure 11). Given the extreme rarity of $z \sim 6$ SMBHs, with
M$_{\rm{BH}} > 10^9$ \msolarc, ($\sim 10^{-9}~{\rm Mpc^{-3}}$), these objects are likely
to have formed in highly biased regions with many more dark matter haloes than the mean number density.
Thus, it seems plausible that nearby clustered sources could provide the necessary flux. This could
be accomplished by, for example, $\sim 5$ sources similar to our primary halo within $2 ~{\rm kpc}$
or one source 5 times brighter.\\
\indent We also note that a radiation background is not the only mechanism which could potentially
delay PopIII star formation. Baryon-dark matter streaming velocities\cite{Tseliakhovich_2010} can
significantly inhibit the collapse of gas in mini-haloes at very high
redshift\cite{Stacy_2011,Fialkov_2012,Naoz_2013} and could reduce the required intensity of the
background field by 2 - 3 orders of magnitude\cite{Tanaka_2014, Visbal_2014b}. In
particular Tanaka \& Li (2014)\cite{Tanaka_2014} have shown that streaming velocities can suppress
molecular cooling all the way up to the ACH limit. 
Likewise, haloes assembling unusually rapidly via successive mergers on time-scales shorter than the 
\molH cooling timescale can avoid star-formation at the minihalo stage\cite{Fernandez_2014}.
We do not include all of these effects in this study, but note that they 
reduce the need for a strong background.
Nonetheless an additional intense nearby source is always required
to prevent \molH forming in the centre of the collapsing halo\cite{Visbal_2014a}.\\
\indent We also point out that even if the abundance of DCBHs formed in the synchronised pair
channel at $z\sim 25$ is very low, the number density at somewhat lower redshift 
could potentially explain the abundance of high-redshift super-massive black holes. Assuming
separations similar to those in our simulations, Visbal et al. (2014)\cite{Visbal_2014a} estimated the
abundance of synchronised pair DCBHs formed between $z = 10-11$ to be $0.0003 ~{\rm Mpc}^{-3}$. If
the probability distribution function of the LW background for these pairs matched that from Ahn et
al. (2009) \cite{Ahn_2009} (see their Figure 11), combining this with the $J_{\rm BG} \gtrsim 100~J_{21}$
background requirement would not reduce the number density below $\sim 10^{-9}~{\rm Mpc^{-3}}$ (which is
similar to the observed number density of high-redshift quasars). It
should be noted that
  this rough estimate is only a lower limit due to the limited box sizes used in both calculations
  cited above and the fact that other large scale effects, as discussed, are neglected.\\

\subsection*{Data Availability\\}
The numerical experiments presented in this work were run with a fork of the enzo code
available from https://bitbucket.org/john\_regan/enzo-3.0-rp, in particular change set d11330f. This altered
version of enzo also requires an altered version of the grackle cooling library, available from
https://bitbucket.org/john\_regan/grackle-cversion, particularly change set d8df240. 

\end{methods}



\begin{addendum}

\item[Materials \& Correspondence] Correspondence and requests for materials
  should be addressed to John Regan.~(email: john.regan@dcu.ie)\nocite{Wise_2009}.
     
 \item  This work was supported by the Science and Technology Facilities Council (grant
numbers ST/L00075X/1 and RF040365) and by NASA grant NNX15AB19G (to ZH). JR acknowledges support 
from the EU commission via the Marie Sk\l{}odowska-Curie Grant - "SMARTSTARS" - grant number 699941.
JW is supported by National Science Foundation grants AST-1333360 and AST-1614333 and Hubble theory 
grants HST-AR-13895 and HST-AR-14326.  PHJ acknowledges the support of the Academy of Finland grant
1274931. This work used the DiRAC Data Centric system at Durham University, operated by the Institute
for  Computational  Cosmology on  behalf  of  the  STFC  DiRAC  HPC  Facility  (www.dirac.ac.uk).
This equipment was funded by BIS National E-infrastructure capital grant ST/K00042X/1, STFC capital
grant ST/H008519/1, and STFC DiRAC Operations grant ST/K003267/1 and Durham University.  DiRAC  is
part  of  the  National  E-Infrastructure. Some of the preliminary numerical simulations were also
performed on facilities hosted by the CSC -IT Center for Science in Espoo, Finland, which are
financed by the Finnish ministry of education. The Center for Computational Astrophysics is supported
by the Simons Foundation. The freely available astrophysical 
analysis code yt\cite{YT} and plotting library matplotlib\cite{matplotlib} was used to construct 
numerous plots within this paper.
\item[Author contributions] JR modified the publicly available Enzo code and
  Grackle codes used in this work, ran and analysed the code results, and wrote the
  initial manuscript. JR, ZH, JW \& EV determined the simulation setup. The radiation particle
  model was conceived and designed by JR, PHJ \& JW. All authors
  contributed to the interpretation of the results, and to the text of the final manuscript.

 \item[Competing Interests] The authors declare that they have no
competing financial interests.

\end{addendum}
\clearpage
\ifpreprint \figureFive \fi


\begin{table*}
\centering
\caption{Supplementary Data Table: Simulation Details}
\vspace{0.2cm}
\begin{tabular}{ | l | c | l | l | l | l | l | l | }
\hline \hline 
\textbf{\em $\rm{Sim\ Name}^{a}$}
& \textbf{\em $\rm{Background}^{b}$}
& \textbf{\em $\rm{z_{on}}^{c}$} & \textbf{\em $z_{collapse}^{d}$} 
& \textbf{\em $T_{\rm sync}^{e}$} & \textbf{\em $T_{\rm on}^{f}$} 
& \textbf{\em $\rm{R_{sep}^{g}}$} & \textbf{\em $\rm{Result}^{h}$} \\
\hline 
z2550\_100\_350 & $100$ & 25.50 & 24.37 & 1.59 & 8.53 & 322 & \textred{Molecular}\\
z2550\_100\_250 & $100$ & 25.50 & 24.19 & 1.59 & 9.97 & 251 & Atomic\\
z2540\_100\_350 & $100$ & 25.40 & 24.40 & 0.87 & 7.51 & 322 & \textred{Molecular}\\
z2540\_100\_250 & $100$ & 25.40 & 24.20 & 0.87 & 9.13 & 276 & Atomic\\
z2530\_100\_350 & $100$ & 25.30 & 24.45 & 0.14 & 6.50 & 321 & \textred{Molecular}\\
z2530\_100\_250 & $100$ & 25.30 & 24.25 & 0.14 & 8.05 & 242 & \textred{Molecular}\\
z25285\_100\_350 & $100$& 25.285 & 24.45 & 0.04 & 6.39 & 321 & Atomic\\
z25285\_100\_250 & $100$& 25.285 & 24.26 & 0.04 & 7,91 & 242 & \textred{Molecular}\\
\hline
z2530\_150\_300 & $150$ & 25.30 & 24.06 & 3.37 & 9.66 & 303 & Atomic\\
z2515\_150\_200 & $150$ & 25.15 & 23.98 & 2.27 & 9.30 & 218 & Atomic\\
z2500\_150\_350 & $150$ & 25.00 & 24.38 & 1.14 & 4.78 & 388 & \textred{Molecular}\\
z2500\_150\_250 & $150$ & 25.00 & 24.10 & 1.14 & 7.10 & 237 & Atomic\\
z2490\_150\_350 & $150$ & 24.90 & 24.41 & 0.38 & 3.83 & 388 & Atomic\\
z2490\_150\_250 & $150$ & 24.90 & 24.18 & 0.38 & 5.68 & 236 & Atomic\\
z2486\_150\_350 & $150$ & 24.86 & 24.46 & 0.1  & 3.12 & 388 & Atomic\\
z2486\_150\_250 & $150$ & 24.86 & 24.23 & 0.07 & 4.88 & 235 & Atomic\\

\hline 
z2500\_200\_250 & $200$ & 24.62 & 24.15 & 3.47 & 6.69 & 390 & \textred{Molecular}\\
z2500\_200\_350 & $200$ & 24.62 & 23.93 & 3.47 & 8.51 & 238 & Atomic\\
z2480\_200\_300 & $200$ & 24.65 & 24.24 & 1.95 & 4.43 & 388 & \textred{Molecular}\\
z2480\_200\_400 & $200$ & 24.65 & 23.99 & 1.95 & 6.48 & 262 & Atomic\\
z2460\_200\_250 & $200$ & 24.60 & 24.33 & 0.40 & 2.15 & 387 & \textred{Molecular}\\
z2460\_200\_400 & $200$ & 24.60 & 24.16 & 0.40 & 3.53 & 235 & Atomic \\
z2457\_200\_250 & $200$ & 24.57 & 24.15 & 0.16 & 3.37 & 360 & Atomic\\
z2457\_200\_400 & $200$ & 24.57 & 23.93 & 0.16 & 5.20 & 262 & Atomic \\
z24555\_200\_250 & $200$ & 24.555 & 24.39 & 0.007 & 1.31 & 387 & \textred{Molecular}\\
z24555\_200\_400 & $200$ & 24.555 & 24.26 & 0.007 & 2.36 & 309 & Atomic\\

\hline
\end{tabular}
\parbox[t]{0.8\textwidth}{The above table contains the simulation name$^a$, 
  the uniform background radiation field in units of J$_{21}^b$, the redshift at which the primary
  halo turns ``on''$^c$, the collapse redshift of the secondary halo$^d$,
  the synchronisation time in Myrs$^e$, the time for which the primary halo was on in Myrs$^f$, 
  the average separation, in parsecs, between the primary and the secondary$^g$ and finally
  the result of the collapse (i.e. an atomic collapse or a molecular collapse$^h$).  All units
  are physical, unless explicitly stated otherwise.}

\label{Table:AllSims}

\end{table*}


\begin{table*}
\centering
\caption{Supplementary Data Table: Radiation Spectrum}
\begin{tabular}{ | l | l | l | l |}
\hline \hline 
\textbf{\em $\rm{Source}$} & \textbf{\em $\rm{L\ [ph/s]}$} &
\textbf{\em $\rm{Energy\ Bins\ [Ev]}$} & \textbf{\em $\rm{Energy\ Fraction}$} \\
\hline 
$\rm{Metal\ Enriched}$ & $1.2 \times 10^{52}$ & $2.0,  12.8,  14.55, 25.05$ & $0.7510, 0.1080, 
1.33 \times 10^{-7}, 2.54 \times 10^{-3}$ \\
$\rm{PopIII}$ & $1.0 \times 10^{52}$ & $2.0,  12.8, 18.37,  47.23$ & $0.4585,  0.1509,  
2.88 \times 10^{-4},  4.77 \times 10^{-2}$ \\
\hline 
\hline

\end{tabular}
\parbox[t]{0.8\textwidth}{The radiation spectra for both the metal enriched galaxy
  model we used in this study as well as a comparable PopIII model. The spectra are convolved with
  a shielding function which is based on a simple
  model of interstellar-medium extinction and accounts for absorption of the ionising radiation
  within the primary halo itself. The density distribution and emission spectra are based on the
  high resolution simulations of Wise \& Cen (2009)$^{72}$\nocite{Wise_2009}. The model convolves the
  spectral energies (above 13.6 eV) with a simple modelling of the optical depth to ionising radiation$^{21}$. 

}
\label{Table:Spectra}
\end{table*}

\end{document}